\documentclass[twoside]{dis07}
\usepackage[latin1]{inputenc}
\usepackage[dvips]{graphicx,epsfig,color}
\usepackage{wrapfig,rotating}
\usepackage{cite}
\usepackage{amssymb,amsmath,array}
\newcommand\SH{\,\mbox{$\sqcup \! \sqcup$}\,}
\newcommand\Li{\,\mbox{${\rm Li}$}\,}
\newcommand\N{\nonumber}

\begin{document}
\title{{\normalsize\sl DESY 07/082     \hfill {\tt arXiv:yymmnnn}
\\ SFB/CPP-07-27 \hfill {   } }\\
Structural Relations between Harmonic Sums up to w=6}

\author{J. Bl\"umlein$^1$ and S. Klein$^1$
%
\thanks{This paper was supported in part by SFB-TR-9: Computergest\"utze 
Theoretische Teilchenphysik and Studienstiftung des Deutschen Volkes.}
%
\vspace{.3cm}\\
%
Deutsches Elektronen-Synchrotron, DESY,
Platanenallee 6, D-15738 Zeuthen, Germany
}

\maketitle

\begin{abstract}
Multiply nested finite harmonic sums $S_{a_1 ... a_n}(N)$ occur in many
single scale higher order calculations in Quantum Field Theory. We discuss 
their algebraic and structural relations to weight {\sf w=6}. As an 
example, we consider the application of these relations to the soft and 
virtual corrections for Bhabha-scattering to $O(\alpha^2)$.
\end{abstract}

\section{Introduction}
Single scale processes in massless Quantum Field Theories \cite{BR1,THRL}, or field 
theories being considered
in the limit $m^2/Q^2 \rightarrow 0$ \cite{HEAV}, both for space- and time-like processes, 
{exhibit} {significant simplifications} {when calculated in} {Mellin space} 
if  compared to representations in momentum-fraction $x-$space. Here, the Wilson coefficients 
and splitting functions are expressed by Nielsen-type integrals 
\begin{equation}
S_{n,p,q}(x) = \frac{(-1)^{n+p+q-1}}{\Gamma(n) p! q!}
\int_0^1 \frac{dz}{z} \ln^{(n-1)}(z) \ln^p(1-zx) \ln^q(1+zx) 
\nonumber
\end{equation}
or harmonic polylogarithms \cite{VR}.
The simplification is, to some extent, due to the structure of 
{Feynman parameter integrals} {which possess a} {Mellin symmetry.}
The respective  expressions can be expressed by finite harmonic sums  $S_{a_1 ... a_n}(N)$ for 
processes to 3--loop 
order \cite{HS1,HS2}, which form the appropriate language. Within the light-cone expansion, or
analogous formalisms for time-like processes, the argument of the harmonic sums are even- or 
odd integers, depending on the process. However, one may consider mathematical generalizations,
continuing the argument analytically to rational, real and complex values  {$N~\epsilon~{\bf Q, 
R, C}$, respectively, \cite{HS1,AC}. In these extensions new relations between the harmonic sums 
are obtained, which lead to more compact representations. Since the hard-scattering cross 
sections usually have to be convoluted with non-perturbative parton densities, it is convenient 
to widely work in Mellin space using analytic representations, also for the solution of the 
evolution equations. This also applies to the treatment of heavy flavor contributions in 
the full phase space, for which concise semi-analytic representations were derived \cite{AB}.  
The final $x-$space results are obtained by a single numerical Mellin--inversion performed 
by a contour integral around the singularities of the problem.

In this note we give a summary on the algebraic and structural relations for finite harmonic 
sums, occurring in hard scattering processes. As an example we consider the virtual and soft 
QED corrections to Bhabha-scattering to $O(\alpha^2)$ in the on-mass-shell scheme \cite{BHABHA} 
to show that also this process
fits to the general basis-representation being derived for various other two-- and three--loop QCD 
processes.
\section{Algebraic Relations}
The complexity of finite harmonic sums is given by $N_{max} = 2 \cdot 3^{w-1}$, with the weight 
$w = \sum_{k=1}^n |a_k|$  growing exponentially. In the $x-$space representation of the 
2--loop Wilson coefficients in QCD, which are described by $w =4$, nearly 80 functions 
emerge, which corresponds to the maximum level possible, cf. \cite{NZ,HS1}.  The algebraic 
relations of finite harmonic sums operate on their index set and are implied by their 
quasi-shuffle algebra \cite{MEH}. The algebraic relations of finite harmonic sums were 
investigated in \cite{ALG} in detail. An example for the shuffle product is
\begin{equation}
S_{a_1,a_2} \SH 
S_{a_3,a_4} =
 S_{a_1,a_2,a_3,a_4}
+S_{a_1,a_3,a_2,a_4}
+S_{a_1,a_2,a_4,a_2} 
+S_{a_3,a_4,a_1,a_2}
+S_{a_3,a_1,a_4,a_2}
+S_{a_3,a_1,a_2,a_4}~.
\nonumber \end{equation}
For a given index set the number of basic harmonic sums is counted by the number of Lyndon words
and can be calculated by a Witt--formula. Investigating the type of harmonic sums emerging in 
physical single--scale problems up to 3--loop order, cf. \cite{BR1,THRL,HEAV}, the index 
$\{-1\}$ never occurs. Their number is $N_{\neg -1} = [(1-\sqrt{2})^w + (1+\sqrt{2})^w]/2$
\cite{DITT}. One therefore may significantly reduce the number of basic functions using $N_{\neg 
-1}$ in the corresponding Witt formula. The following table illustrates the corresponding 
complexities of combinations $\#_c$ and the number of algebraic basis elements $\#_b$ in 
dependence of the weight $w$. The initial complexity in case of the absence of indices $a_l = 
-1$ is lower than the number of algebraic basis elements in the complete case.

\vspace{2mm}\hspace{3cm}
\begin{tabular}{|c|r|r|r|r|r|r|}
\hline
$w$    &  1 &  2  &  3 &  4 &   5 &   6 \\
\hline
$\#_c$ &  2 &  8  & 26 & 80 & 242 & 728 \\
$\#_b$ &  2 &  5  &  13 & 31 &  79 & 195 \\
\hline
$\#_c(\neg -1)$ &  1 &  4  & 11 & 28 &  69 & 168 \\
$\#_b(\neg -1)$ &  1 &  3  &  7 & 14 &  30 &  60 \\
\hline
\end{tabular}

\section{Structural Relations}
The algebraic relations lead to a first reduction of the complexity of finite harmonic 
sums. They are independent of the value of these quantities. Beyond these relations, the
{\sf structural relations} are of a more specific type, cf.\cite{HS1,JB04}. There are three types
of these relations. The first class is implied by allowing half-integer values for $N$. A second 
class emerges through partial integration using the representations of harmonic sums through
Mellin-transforms e.g. of weighted harmonic polylogarithms. The third set is implied by 
differentiating harmonic sums w.r.t. their argument, which requires $N~\epsilon~{\bf R}$. To 
illustrate case~1 we represent 
\begin{equation}
\frac{1}{1-x^2} = \frac{1}{2} \left[ \frac{1}{1-x} + \frac{1}{1+x}\right]. \nonumber
\end{equation}
The Mellin transform of this equation implies, that $S_{-1}(N)$ is linearly dependent of
$S_1(N)$, if $N~\epsilon~{\bf Q}$. Various relations of this type emerge at higher weight 
\cite{JB04}. If $N~\epsilon~{\bf R}$, one may differentiate harmonic sums and obtains
\begin{equation}
\frac{d}{dN} S_{a_1 ... a_n}(N) = \int_0^1~dx~x^{N-1} \ln(x) f(x)~.
\nonumber
\end{equation}
$S_{a_1 ... a_n}(N)$ is given by the Mellin transform of $f(x)$. One may represent the 
derivative in terms of polynomials of harmonic sums and their values at $N \rightarrow \infty$.
The analytic continuation of a harmonic sum to $N~\epsilon~{\bf C}$ is a meromorphic function 
with poles at the non-positive integers. Up to polynomials growing $\propto S_1^m(N),~~|N| 
\rightarrow \infty $, which can 
be separated, harmonic sums are factorial series. I.e. they obey an analytic recursion relation 
$F(z+1) \rightarrow F(z)$ and one may calculate their asymptotic representation for $z \rightarrow  
\infty$ analytically. In this way these functions are uniquely defined in the complex plane.
Due to the above properties  we now define equivalence classes of representations, which contain 
a harmonic sum and all its derivatives. We will only count these equivalence classes, since all 
derivatives can be easily found analytically if the respective lowest weight sum of the class is 
known. In the following table we summarize the functions of the basic Mellin transforms up to 
$w = 5$. The case
$w = 6$ is still to be completed.

{
\begin{equation}
\begin{array}{llll}
\hspace*{-1cm}
w=1 & 1/(x-1)_+           &                       &      \\
\hspace*{-1cm}
w=2 & \ln(1+x)/(x+1)      &                       &      \\        
\hspace*{-1cm}
w=3 & \Li_2(x)/(x \pm 1)  &                       &      \\         
\hspace*{-1cm}
w=4 & \Li_3(x)/(x + 1)    & S_{1,2}(x)/(x \pm 1)  &      \\
\hspace*{-1cm}
w=5 & \Li_4(x)/(x \pm 1)  & S_{1,3}(x)/(x + 1)    & S_{2,2}(x)/(x \pm 1) \\
      & \Li_2^2(x)/(x+1)  & [S_{2,2}(-x) - \Li_2^2(-x)/2]/(x \pm 1) &    \\
\hspace*{-1cm}
w=6 & \Li_5(x)/(x+1)        & S_{1,4}(x)/(x \pm 1) & S_{2,3}(x)/(x \pm 1) \\
\hspace*{-1cm}
    & S_{3,2}(x)/(x \pm 1) & \ldots                     &  
\end{array}\nonumber\end{equation}
}

\vspace{1mm} \noindent
These functions emerge in the following processes~:~~~~~~~~number of fct. 
\vspace{1mm} \noindent

\begin{tabular}{lll}
{$\bullet~~O(\alpha)~~w \leq 2$ }  & {Wilson Coefficients/anom. dim.}  &  \hspace{5mm} 
{$\#1$}\\
{$\bullet~~O(\alpha^2)~ w \leq 3$} & {Anomalous Dimensions} &  \hspace{5mm} {$\#2$}\\
{$\bullet~~O(\alpha^2)~ w \leq 4$} & {Wilson Coefficients}  &  \hspace{5mm} {$\# \leq 
5$}\\
{$\bullet~~O(\alpha^3)~ w \leq 5$} & {Anomalous Dimensions} &  \hspace{5mm} {$\# 15$}\\
{$\bullet~~O(\alpha^3)~ w \leq 6$} & {Wilson Coefficients}  &  
\hspace{5mm} {$\# 29+$}\\
\end{tabular}

\section{Example: Bhabha Scattering}
As an example we consider the soft- and virtual corrections to Bhabha-scattering up to
$O(\alpha^2)$, $T_2(x) = \delta_0^{(2)}(x) \times (1-x+x^2)/x^2$, \cite{BHABHA}. 
The $x-$space expression simplifies in Mellin $N$ space 
and depends on {\it three} basic functions $S_{2,1,1}, S_{2,1}$ and $S_1$ only aside of 
polynomial pre-factors~: 

{\footnotesize
  \begin{eqnarray}   
   T_2(N) &=&  
   \frac{248+15\,{N}^{2}+{N}^{4}}
   {2(N-2)(N-1)N(N+1)(N+2)}
   { S_{1,1,1,1}}(N)
   -\frac {2} 
   {(N-1)(N+1)}
   {\bf S_{2,1,1}} (N)  
   \N\\ & &
   +\frac{-340+120\,N+17\,{N}^{2}+18\,{N}^{3}-31\,{N}^{4}}
   {2(N-2)(N-1)N(N+1)(N+2)}
    { S_{3,1}} (N) 
   -
   \frac {-304 + 278 N + 81 N^2 - 38 N^3 + 19 N^4}{8 (N-2) (N-1) N (N+1) (N+2)}
   { S_4}(N)
   \N\\ & &
   +\frac {304-328\,N-500\,{N}^{2}+330\,{N}^{3}-6\,{N}^{4}+6\,{N}^{5}-2
   \,{N}^{6}+4\,{N}^{7}}
   {(N-2)^{2}(N-1)^{2}{N}^{2}(N+1)(N+2)}
   {\bf S_{2,1}} (N)
   \N\\ & &
   +\frac{-112-4\,{N}^{2}-4\,{N}^{4}}
   {(N-2)(N-1)N(N+1)(N+2)}
   { S_{2,1}}(N){\bf S_1}(N)
   +\frac{-48+8\,N+6\,{N}^{2}+7\,{N}^{3}}
   {(N-1)N(N+1)(N+2)}
   { S_3}(N) { S_1} (N) 
   \N\\ & &
   +\frac{-1840+292\,N+5532\,{N}^{2}+827\,{N}^{3}-1978\,{N}^{4}-274\,{N}^{5}
   +36\,{N}^{6}+19\,{N}^{7}-22\,{N}^{8}}
   {4(N-2)^{2}(N-1)^{2}{N}^{2}(N+1)^{2}(N+2)}
   { S_{1,1,1}}(N)
   \N
\end{eqnarray}\begin{eqnarray}
& &
   +\frac{128-56\,N-252\,{N}^{2}+54\,{N}^{3}+177\,{N}^{4}-91\,{N}^{5}
   +19\,{N}^{6}+9\,{N}^{7}}
   {2(N-2)(N-1)^{2}{N}^{2}(N+1)^{2}(N+2)}
   { S_3}(N) 
   \N
\\ 
& &
   +\frac{4032-2048\,N-14200\,{N}^{2}+5036\,{N}^{3}+23610
   \,{N}^{4}+2521\,{N}^{5}-12342\,{N}^{6}}
   {4(N-2)^{3}(N-1)^{3}N^3(N+1)^{3}(N+2)}
   { S_{1,1}}(N)   
   \N\\ & &
   +\frac{-3365\,{N}^{7}+2148\,{N}^{8}+903
   \,{N}^{9}+14\,{N}^{10}-167\,{N}^{11}+50\,{N}^{12}}
   {4(N-2)^{3}(N-1)^{3}N^3(N+1)^{3}(N+2)}
   { S_{1,1}}(N)
   \N\\ & &
   +\frac{-124+16\,N+24\,{N}^{2}-4\,{N}^{3}-14\,{N}^{4}}
   {(N-2)(N-1)N(N+1)(N+2)}
   { S_{1,1}}(N) { \zeta(2)}
   +\frac{424-118\,N+9\,{N}^{2}-2\,{N}^{3}+23\,{N}^{4}}
   {4(N-2)(N-1)N(N+1)(N+2)}
   { S_2}(N){ S_{1,1}}(N)
   \N\\ & &
   +\frac{224+144\,N-1216\,{N}^{2}-56\,{N}^{3}+1786\,{N}^{4}+641\,{N}^{5}
   -406\,{N}^{6}}
   {4(N-2)^{2}(N-1)^{3}N^3(N+1)^3(N+2)}
   { S_2}(N)
   \N\\ & &
   +\frac{17\,{N}^{7}-308\,{N}^{8}+141\,{N}^{9}-56\,{N}^{10}+{N}^{11}}
   {4(N-2)^{2}(N-1)^{3}N^3(N+1)^3(N+2)}
   { S_2}(N)
   +\frac{58+21\,N+{N}^{2}+15\,{N}^{3}+10\,{N}^{4}}
   {(N-2)(N-1)N(N+1)(N+2)}
   { S_2}(N){ \zeta(2)} 
   \N\\ & &
   +\frac{232-384\,{N}^{2}-17\,{N}^{3}+286\,{N}^{4}-128\,{N}^{5}-14\,{N}^{6}
   +{N}^{7}}
   {4(N-2)(N-1)^{2}{N}^{2}(N+1)^{2}(N+2)}
   { S_2}(N){ S_1}(N)  
   \N\\ & &
   +\frac {-560-26\,N-31\,{N}^{2}-10\,{N}^{3}-33\,{N}^{4}}
   {8(N-2)(N-1)N(N+1)(N+2)}
   { S_2}(N)^2
   \N\\ & &
   +\frac{576+1088\,N-3280\,{N}^{2}-5136\,{N}^{3}+11764\,{N}^{4}+20392\,{N}^{5}
   -17385\,{N}^{6}-30114\,{N}^{7}}
   {4(N-2)^{3}(N-1)^{4}{N}^{4}(N+1)^{4}(N+2)}
   { S_1}(N)
   \N\\ & &
   +\frac{5984\,{N}^{8}+17228\,{N}^{9}-1228\,{N}^{10}-2754\,{N}^{11}
   -112\,{N}^{12}-8\,{N}^{13}+33\,{N}^{14}-24\,{N}^{15}}
   {4(N-2)^{3}(N-1)^{4}{N}^{4}(N+1)^{4}(N+2)}
   { S_1}(N)
   \N\\ & &
   +\frac{-56+336\,N+522\,{N}^{2}+424\,{N}^{3}-53\,{N}^{4}-500\,{N}^{5}
   +60\,{N}^{6}+28\,{N}^{7}-5\,{N}^{8}}
   {2(N-2)^{2}(N-1)^{2}{N}^{2}(N+1)^{2}(N+2)}
   { S_1}(N){ \zeta(2)}
   \N\\ & &
   +\frac {64+6\,{N}^{2}+{N}^{3}}
   {(N-2)(N-1)N(N+1)}
   { S_1}(N)\zeta(3)
   +\frac{2112+608\,N+76\,{N}^{2}-140\,{N}^{3}+107\,{N}^{4}}
   {10(N-2)(N-1)N(N+1)(N+2)}
   {{ \zeta(2)}}^{2}
   \N\\ & &
   +\frac {-224-136\,N+1688\,{N}^{2}+1290\,{N}^{3}-1998\,{N
   }^{4}-1997\,{N}^{5}+198\,{N}^{6}}
   {2(N-2)^2(N-1)^{3}{N}^{3}(N+1)^{3}(N+2)}
   { \zeta(2)}
   \N\\ & &
   +\frac{405\,{N}^{7}+376\,{N}^{8}-119\,{N}^{9
   }+56\,{N}^{10}+5\,{N}^{11}}
   {2(N-2)^2(N-1)^{3}{N}^{3}(N+1)^{3}(N+2)}
   { \zeta(2)}
   \N\\ & &
   +\frac{-552+144\,N+1654\,{N}^{2}-370\,{N}^{3}-361\,{N}^{4}+19\,{N}^{5}
   +35\,{N}^{6}-25\,{N}^{7}}
   {2(N-2)^{2}(N-1)^{2}{N}^{2}(N+1)^{2}}
   \zeta(3) 
   \N\\ & &
   +\frac{P_1(N)}
   {16(N-2)^3(N-1)^{5}(N+1)^{5}{N}^{5}(N+2)}
   \N\\ 
%
%
& &   +4\frac{{N}^{4}-{N}^{2}+12}
   {(N-2)(N-1)N(N+1)(N+2)}
   f_{0,2}
   -2\frac {{N}^{4}-{N}^{2}+12}
   {(N-2)(N-1)N(N+1)(N+2)}
   f_{0,1}^2
   \N
\end{eqnarray}
}
with 
{\footnotesize
\begin{eqnarray}
P_1 &=&  320-64\,N-1920\,{N}^{2}+1600\,{N}^{3}+6524\,{N}^{4}-
   14872\,{N}^{5}-19036\,{N}^{6}+31543\,{N}^{7}
\nonumber\\ & &
-43960\,{N}^{8}-13935\,{N}
   ^{9} + 65372\,{N}^{10}+26822\,{N}^{11}-44576\,{N}^{12}-9558\,{N}^{13}
\nonumber\\ & &
+
   9840\,{N}^{14}+339\,{N}^{15}+428\,{N}^{16}
-371\,{N}^{17}+128\,{N}^{18}
\nonumber \end{eqnarray}}
There are no alternating sums contributing, unlike the case for 
other 2--loop Wilson coefficients in QCD \cite{BR1,HEAV}, where generally up to 6 functions
emerge.
\section{Conclusions}
The single-scale quantities in Quantum 
Field Theories to 3 Loop Order, corresponding to $w = 6$, 
{can be represented in a 
polynomial ring spanned by} {a few Mellin transforms} {of  the 
above} {basic functions,} {which are the same for all known 
processes, including QED processes, as the soft- and virtual corrections to Bhabha-scattering. 
This points to their general nature.}
{The} {basic Mellin transforms} {are} 
\textcolor{black}{meromorphic functions} {with single poles at the 
non-positive integers.} 
{The total amount of harmonic sums reduces due to} 
{algebraic relations} \textcolor{black} as a consequence of the {index structure},
{and} {structural relations} continuing the argument to \textcolor{black}{ 
N$~\epsilon~{\bf Q}$, N$~\epsilon~{\bf R}$}.
{They can be represented in terms of} {factorial 
series} {up to simple ``soft components''. This allows an exact}
{analytic continuation.}
{Up to} {$w = 6$} {physical (pseudo-) observables 
are free of harmonic sums with} {index =$ \{-1\}$}. {To} 
{$w = 5$} {all numerator functions are} {Nielsen 
integrals.}
%
%
%

\begin{footnotesize}

\end{footnotesize}
\end{document}